\documentclass[12pt]{iopart}

\usepackage{graphicx}
\usepackage{caption}
\newcommand {\PT}          {\ensuremath{p_T}}
\newcommand {\ET}          {\ensuremath{E_T}}

\newcommand{\RAA}        {\ensuremath{R_{AA}}}

\newcommand{\Npart}      {\ensuremath{N_{part}}}

\begin{document}

\title[ ]{Overview of experimental results in PbPb collisions 
at $\sqrt{s_{_{\mathrm{NN}}}}$ = 2.76 TeV by the CMS Collaboration}

\author{Bolek Wyslouch for the CMS collaboration}

\address{LLR-\'Ecole Polytechnique/IN2P3, Palaiseau, France and Massachusetts Institute of Technology,
Cambridge, MA, USA}
\ead{wyslouch@mit.edu}
\begin{abstract}
The CMS experiment at the LHC is a general-purpose apparatus with a set of large acceptance and high granularity detectors for hadrons, electrons, photons and muons, providing unique capabilities for both proton-proton and ion-ion collisions. The data collected during the November 2010 PbPb run at $\sqrt{s_{_{\mathrm{NN}}}}$ = 2.76 TeV was analyzed and multiple measurements of the properties of the hot and dense matter were obtained. Global event properties, detailed study of jet production and jet properties, isolated photons, quarkonia and weak bosons were measured and compared to pp data and Monte Carlo simulations.  
\end{abstract}

The LHC first produced collisions of heavy-ions in November of 2010. Lead ions were accelerated to a nucleon-nucleon center of mass energy of 
$\sqrt{s_{_{\mathrm{NN}}}}$ = 2.76~TeV, a factor of roughly 14 times higher than previously achieved in laboratory experiments. 
Nuclear interactions at these energies were expected to produce a hot and dense system with a significantly higher temperature and a
significantly longer duration than at RHIC. The CMS detector was used to study the properties of the matter created in this new regime.
CMS is a general purpose particle detector very well suited to study high energy nuclear collisions\cite{JINST}. High precision tracking and calorimetry with fine granularity are augmented with a sophisticated multi-level triggering system. The initial analysis of CMS heavy-ion data revealed
information ranging from global particle production, which is related to the initial state formed in the collision, and various observables sensitive
to the hydrodynamical expansion of the system, to detailed probes of the produced medium by studying the propagation of a broad range of particles.
More details on the
analysis of specific results can be found in the proceeding on global properties of charged particle production\cite{QM11_Krajczar}, high multiplicity\cite{QM11_Velicanu,ridgeppCMS} and HBT\cite{QM11_Padula} in pp collisions, elliptic flow \cite{QM11_Velkovska,QM11_Zhukova}, dihadron fluctuations \cite{QM11_Li, QM11_Callner,correlationscentralCMS}, charged particle $\RAA$ \cite{QM11_Yoon,QM11_Lee}, jet quenching and fragmentation \cite{QM11_CRoland,QM11_Tonjes,QM11_Yilmaz,QM11_Nguyen,jetquenchingCMS}, photons \cite{QM11_Kim}, quarkonia suppression
\cite{QM11_Silvestre, QM11_Dahms, QM11_Hu, QM11_Jo, upsilonsuppressionCMS} and weak bosons \cite{QM11_Robles,ZHICMS}.

The CMS detector incorporates a broad suite of high precision subsystems which were originally optimized for very high energy pp collisions
but which also provide unprecedented capabilities for studying nuclear collisions. The inner tracker consists of 3 layers of silicon
pixel detectors placed close to the interaction region followed by 10 layers of single- or double-sided 
silicon micro strip detectors. This inner tracker is surrounded by the electromagnetic calorimeter (ECAL)
which uses lead tungstate $(\rm{PbWO}_4)$ crystals with coverage in pseudorapidity up to $|\eta|<3$. 
The ECAL is surrounded by a brass/scintillator sampling hadron calorimeter (HCAL) with the same $\eta$ coverage.
Additional hadronic calorimetry coverage up to a pseudorapidity of $|\eta|<5.2$ is provided by an iron/quartz-fiber calorimeter (HF).
The inner tracker and the calorimeters are installed inside a 13~m long, 6~m-inner-diameter, 3.8~T superconducting solenoid.
The return field is large enough to saturate 1.5~m of iron, allowing 4 layers of muon stations located outside the solenoid. 
Each muon station consists of several layers of aluminum drift tubes (DT)
in the barrel region surrounding the solenoid and cathode strip chambers (CSC) in the endcap region, complemented by
resistive plate chambers (RPC). The very forward region is equipped with the CASTOR calorimeter covering 
$-6.7<\eta<-5.2$ and a pair of Zero-Degree Calorimeters (ZDC) situated at about $\pm 140$~m from the interaction point. All of these
detectors have sufficient granularity and resolution to function well even in the highest multiplicities encountered in PbPb collisions.
The CMS trigger and data acquisition systems were specially configured for the 2010 PbPb run to prepare for possible surprises related to the large multiplicities expected in these collisions. In particular, the silicon micro strip tracker and the ECAL and HCAL calorimeters were read out in non-zero suppressed mode, recording information from all channels. The zero suppression was done during the offline processing using algorithms optimized for PbPb environment. The maximum inelastic PbPb collision rate achieved during the run was about 220~Hz while the maximum rate that could be written to tape was about 140~Hz. CMS employed its flexible triggering system to select and write to tape all events containing high $\PT$ jets, photons and muons while recording a scaled-down random sample of minimum bias events. 
For analysis of PbPb events, it is important to determine the overlap or impact parameter of the
two colliding nuclei, usually called ``centrality''. Centrality in CMS was determined using the total sum of
transverse energy in reconstructed towers from both positive and negative HF calorimeters covering $2.9<|\eta|<5.2$. For some analyses
the forward energy carried by neutral spectator fragments measured in the ZDC was used as a cross-check of centrality determination.
Centrality for specific event classes is expressed as a percentage of the inelastic nucleus-nucleus interaction cross section.

\begin{figure}[ht!]
\begin{center}
   	\includegraphics[width=0.45\textwidth]{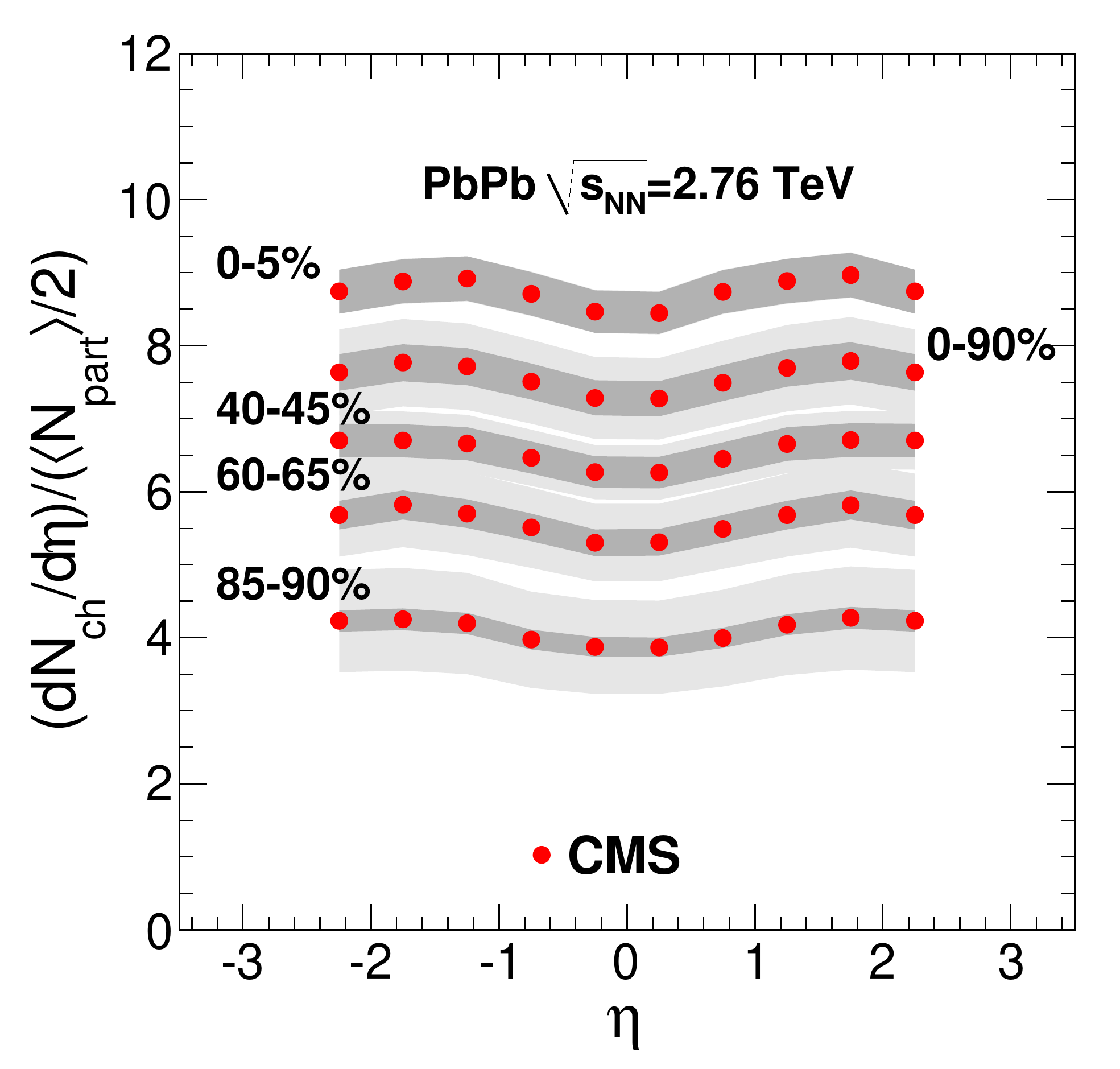}
       \hspace{6mm}
       \includegraphics[width=0.45\textwidth]{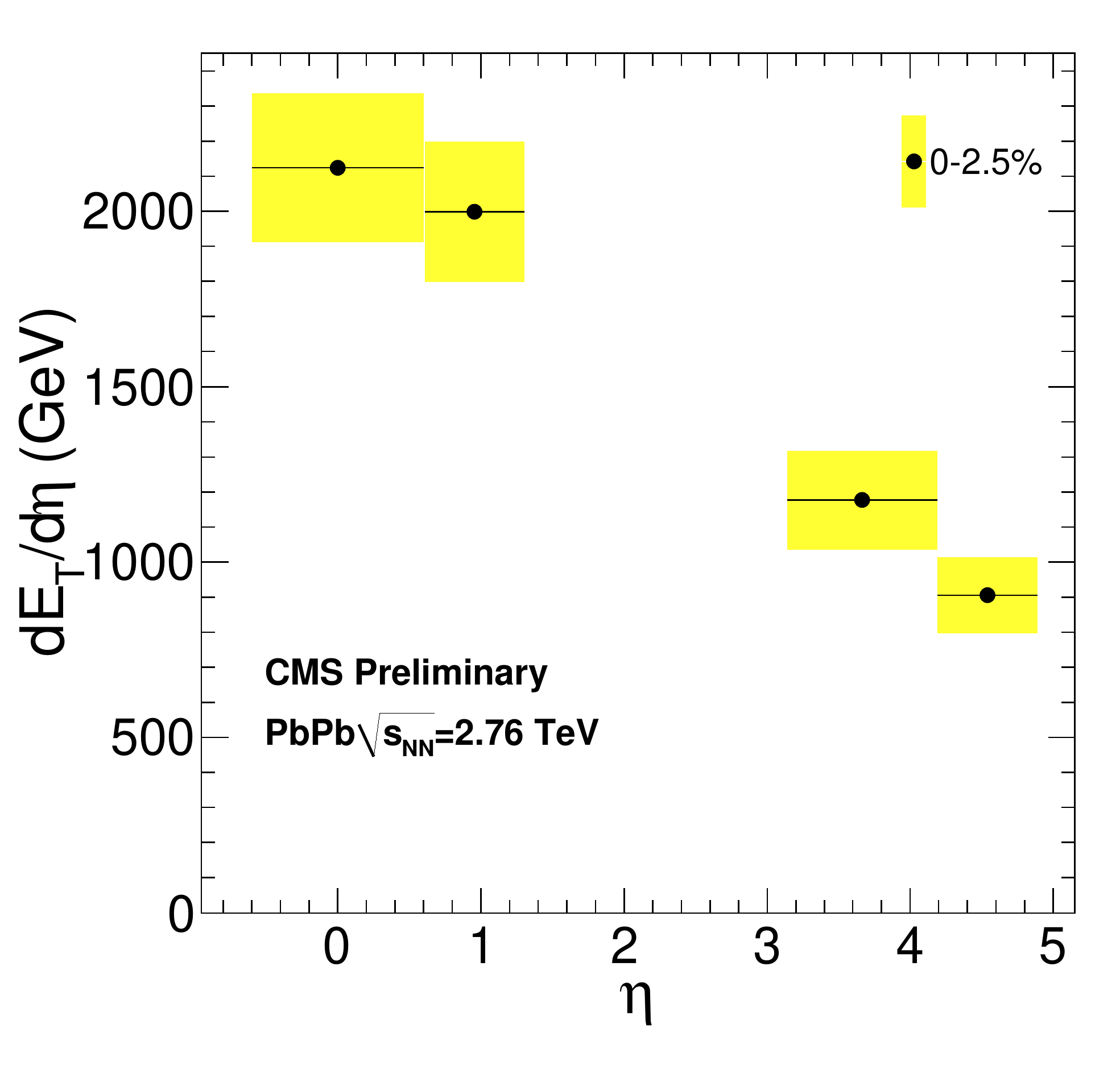} 
       \caption{ 
       (left) Charged particle multiplicity density $dN_{ch}/{d\eta}$ per participant pair for different centralities as a function of $\eta$. (right) Transverse energy density $dE_{T}/{d\eta}$ as a function of $|\eta|$ for most central events (0-2.5\%)}
\label{fig:multiplicity_et}
\end{center}
\end{figure}

The multiplicity of charged particles produced in the central rapidity region is a key observable
to characterize the properties of the matter created in heavy-ion collisions. The measurement of charged particle multiplicity density $dN_{ch}/{d\eta}$ as a function of collision geometry is shown in Fig.~\ref{fig:multiplicity_et}(left). The data was taken without a magnetic field 
in order to include particles with transverse momenta
down to about 30~MeV/c. The number of charged hadrons was obtained by two methods based on the inner silicon pixel system. One technique involved counting the number of reconstructed single particle hits in the pixel detector, while the other formed hit pairs (`tracklets') from the different detector layers.

An observable that is sensitive to the energy density achieved in the ion-ion collisions is the transverse energy $E_T$. Together with multiplicity measurement it gives information of how the initial energy is converted into particle production. 
The broad array of CMS calorimeters was used to measure $dE_{T}/d\eta$ over a wide range of $\eta$. This transverse energy density for the most central PbPb collisions is over 2~TeV per unit pseudorapidity at $\eta=0$ as shown in Fig.~\ref{fig:multiplicity_et}(right). This is about three times higher than at RHIC. 

\begin{figure}[ht!]
\begin{center}
   	\includegraphics[width=0.39\textwidth]{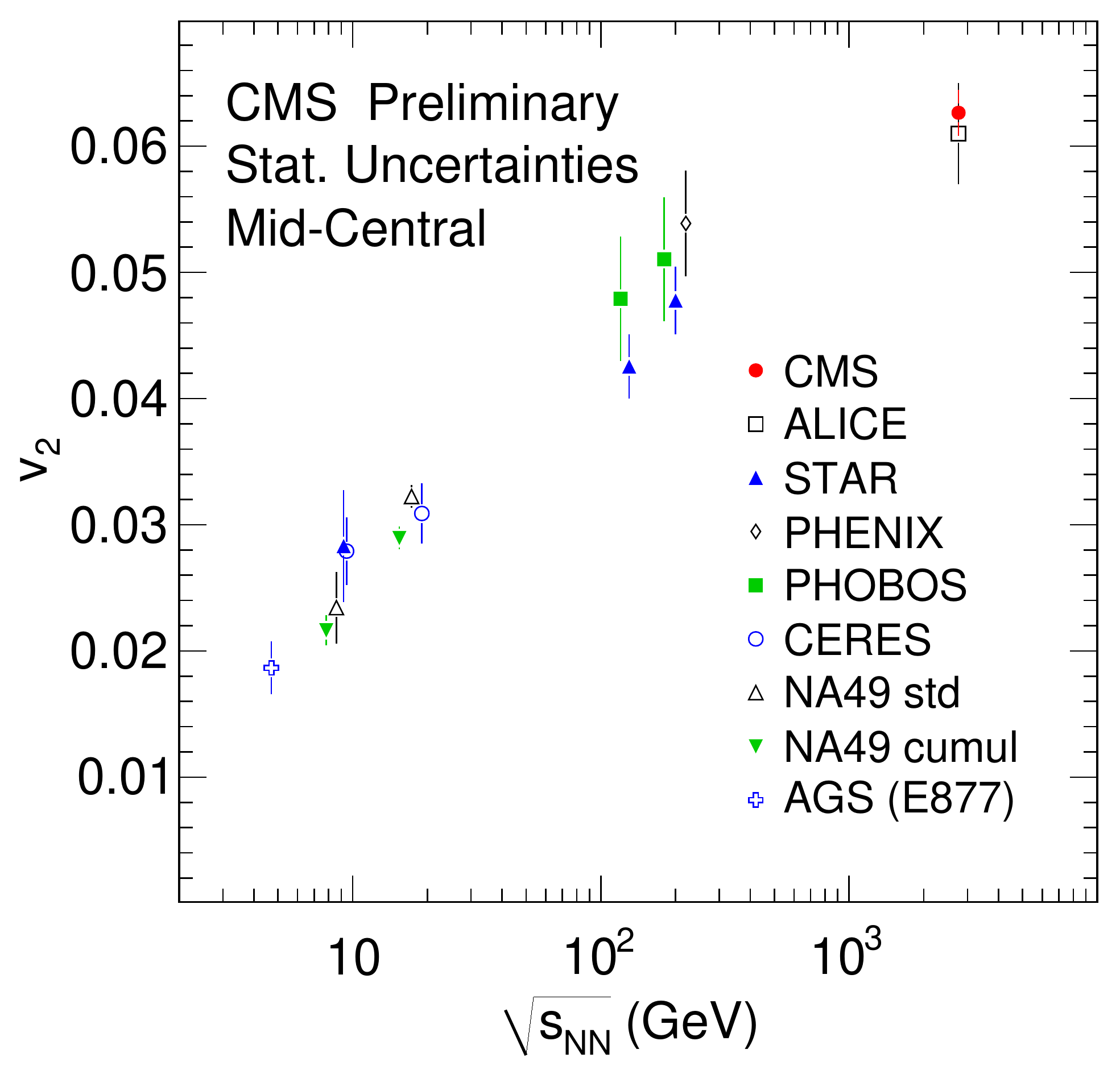}
       \hspace{6mm}
       \includegraphics[width=0.48\textwidth]{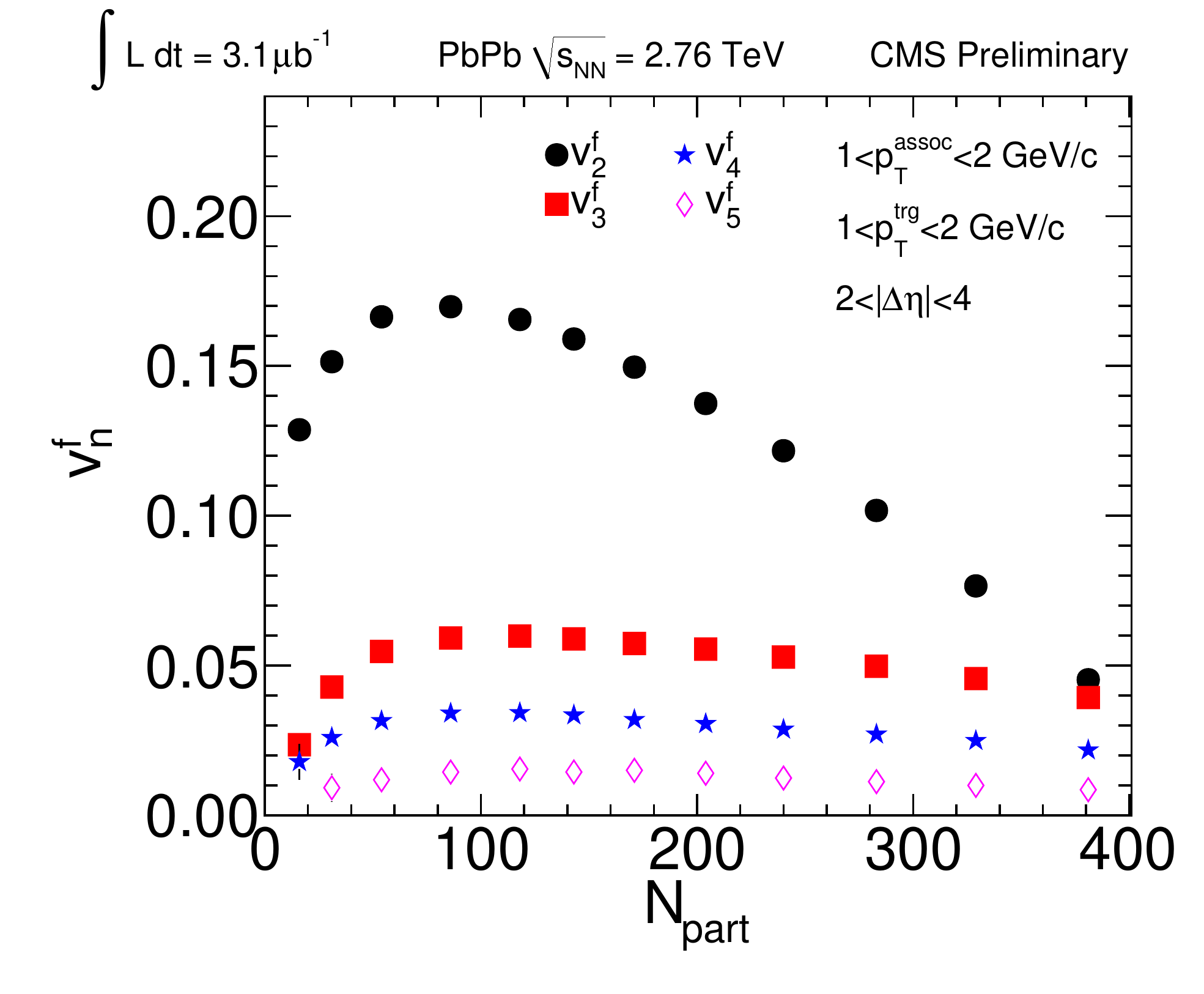}
       \caption{ 
       (left) Integrated flow coefficient $v_2\{LYZ\}$ compared to measurements from other experiments
        for mid-central collisions (20-30\%) 
       (right) Higher order Fourier coefficients as a function of centrality extracted using dihadron correlations}
\label{fig:flow}
\end{center}
\end{figure}

The hydrodynamic expansion of matter produced in peripheral heavy-ion collisions as well as the fluctuation of initial state lead to an azimuthal anisotropy in particle production. This anisotropy was studied using several methods, including event plane, 2- and 4-particle cumulants, Lee-Yang Zeros (LYZ) as well as dihadron correlations in the full CMS acceptance.
The effect is strongest for semi-peripheral events and in the flow coefficient $v_2$.  Detailed measurements of the $v_2$ parameter and the higher harmonics (up to 6th order) as a function of centrality, $\PT$ and pseudorapidity were conducted using charged tracks reconstructed in the silicon tracker. The integrated $v_2$ extrapolated to $\PT=0$ as measured by CMS using the LYZ method for events in a centrality bin of $20-30\%$ is compared to the measurements at lower energies in Fig.~\ref{fig:flow}(left). The relatively small increase compared to RHIC indicates that the hydrodynamic properties are similar. Higher order flow coefficients were also measured as a function of centrality using dihadron correlations and are shown in Fig.~\ref{fig:flow}(right). Note that the higher order coefficients change little with centrality and remain important even for the most central collisions.

\begin{figure}[ht!]
\begin{center}
   	\includegraphics[width=0.4\textwidth]{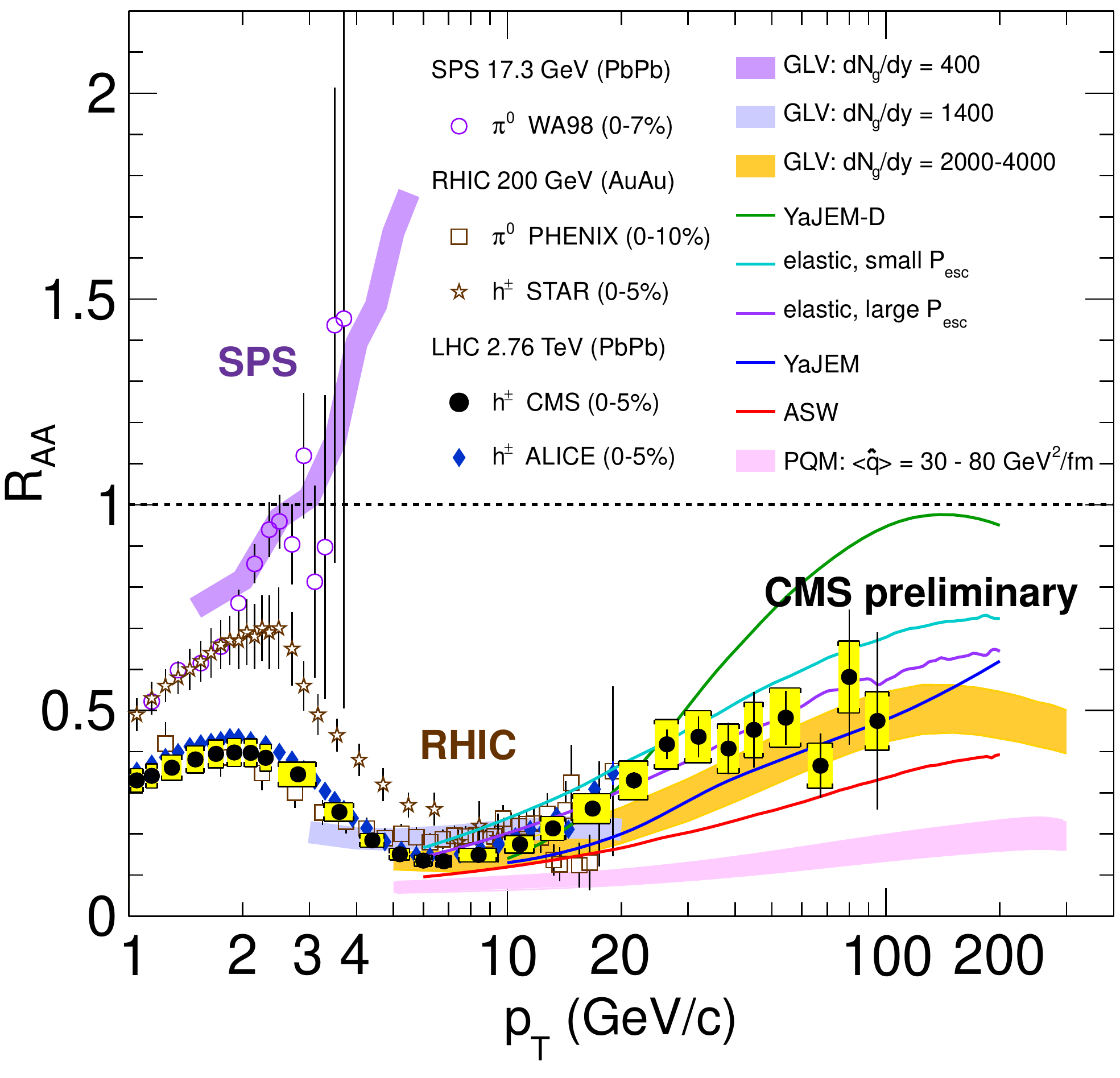}
       \hspace{6mm}
       \includegraphics[width=0.4\textwidth]{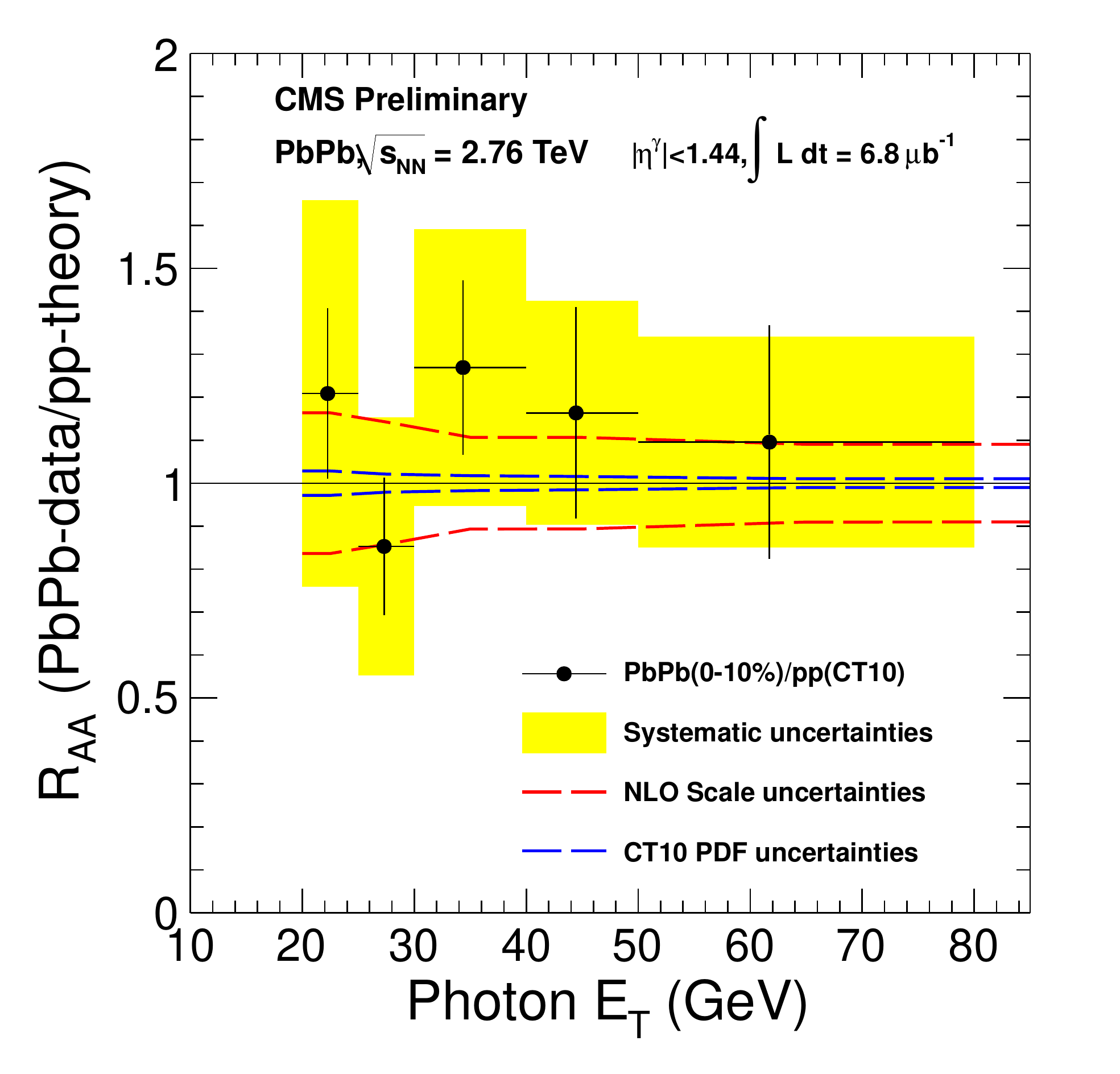} 
       \caption{
              (left) Comparison of data for $\RAA$ as a function of $\PT$ for neutral pions and charged hadrons in central heavy-ion collisions at three different center-of-mass energies and several theoretical predictions~\cite{QM11_Yoon}. (right) $\RAA$ of isolated photons as a function of $\PT$ for central events (0-10\%).} 
\label{fig:raa}
\end{center}
\end{figure}

The modification of charged particle $\PT$ spectrum, compared to nucleon-nucleon collisions at the same energy can shed light on the detailed mechanism by
which hard partons lose energy traversing the medium. The nuclear modification ratio $R_{AA}$ was measured in CMS for
all charged particles with $\PT$ up to 100~GeV/c. The ratio is plotted in Fig.~\ref{fig:raa}(left) and
compared to theoretical predictions.

High transverse energy ($\ET$) prompt photons in nucleus-nucleus collisions 
are produced directly from the hard scattering of two partons. Once produced, photons traverse the
hot and dense medium without any interaction, hence they provide a direct test of perturbative QCD
and the nuclear parton densities. However, the measurement of prompt photons is complicated by the much larger
background coming from the electromagnetic decays of neutral mesons
produced in the fragmentation of other hard scattered partons. 
One can suppresses a large fraction of these decay photon backgrounds by imposing isolation cuts
on the reconstructed photon candidates. A measurement of direct photon production as a function
of centrality and $\PT$ is compared to the NLO pQCD predictions \cite{QM11_Kim}. The ratio $R_{AA}$ as a function of photon $\PT$ for the 
most central events is shown in Fig.~\ref{fig:raa}(right). Within statistical uncertainties, the direct photons are not suppressed as compared to pp collisions.

\begin{figure}[ht!]
\begin{center}
   	\includegraphics[width=0.9\textwidth]{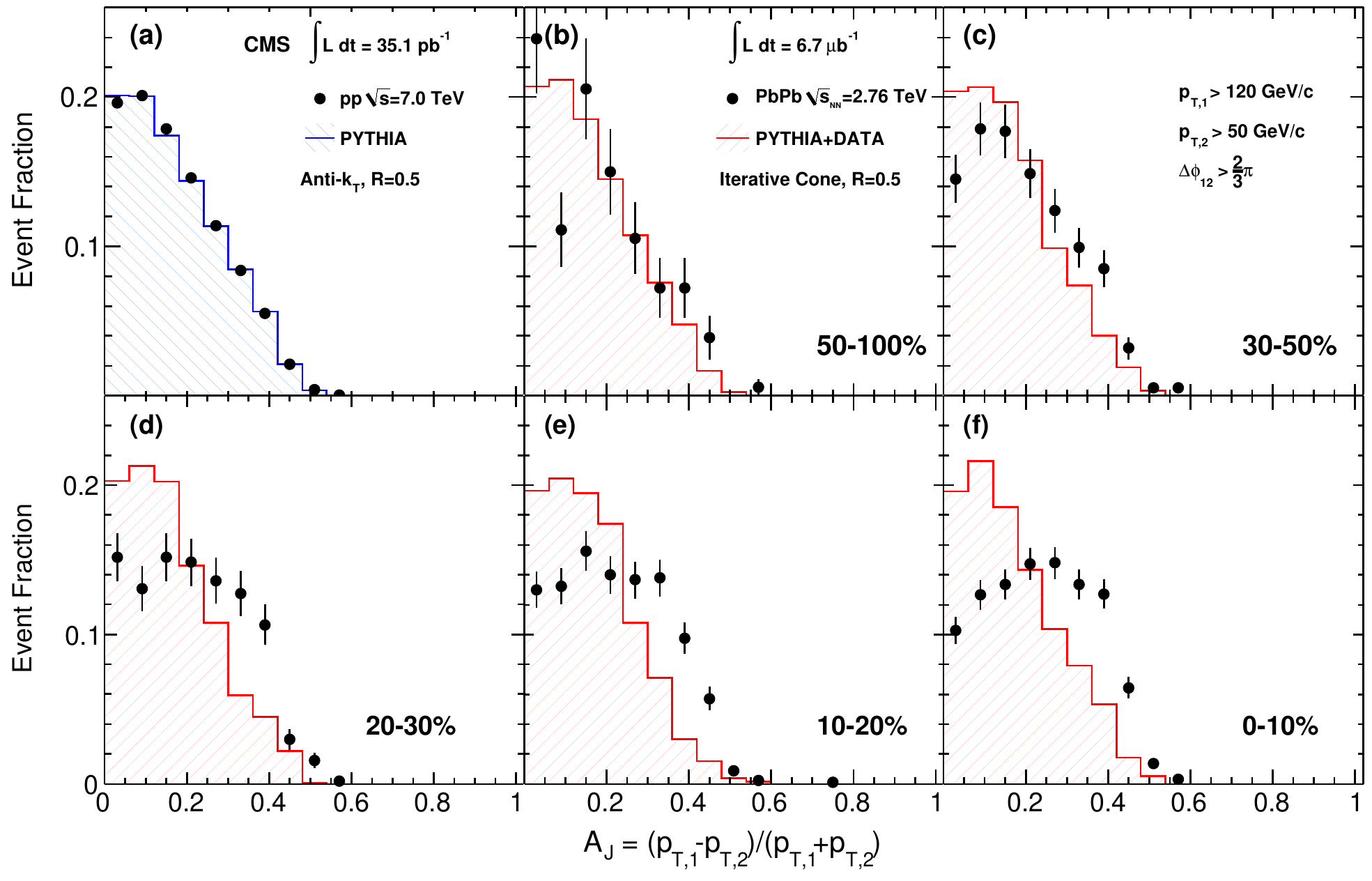}
       \caption{Calorimetric jet imbalance in dijet events as a function of collision centrality for pp and PbPb events.}
\label{fig:dijet_imbalance}
\end{center}
\end{figure}

\begin{figure}[ht!]
\begin{center}
   	\includegraphics[width=0.9\textwidth]{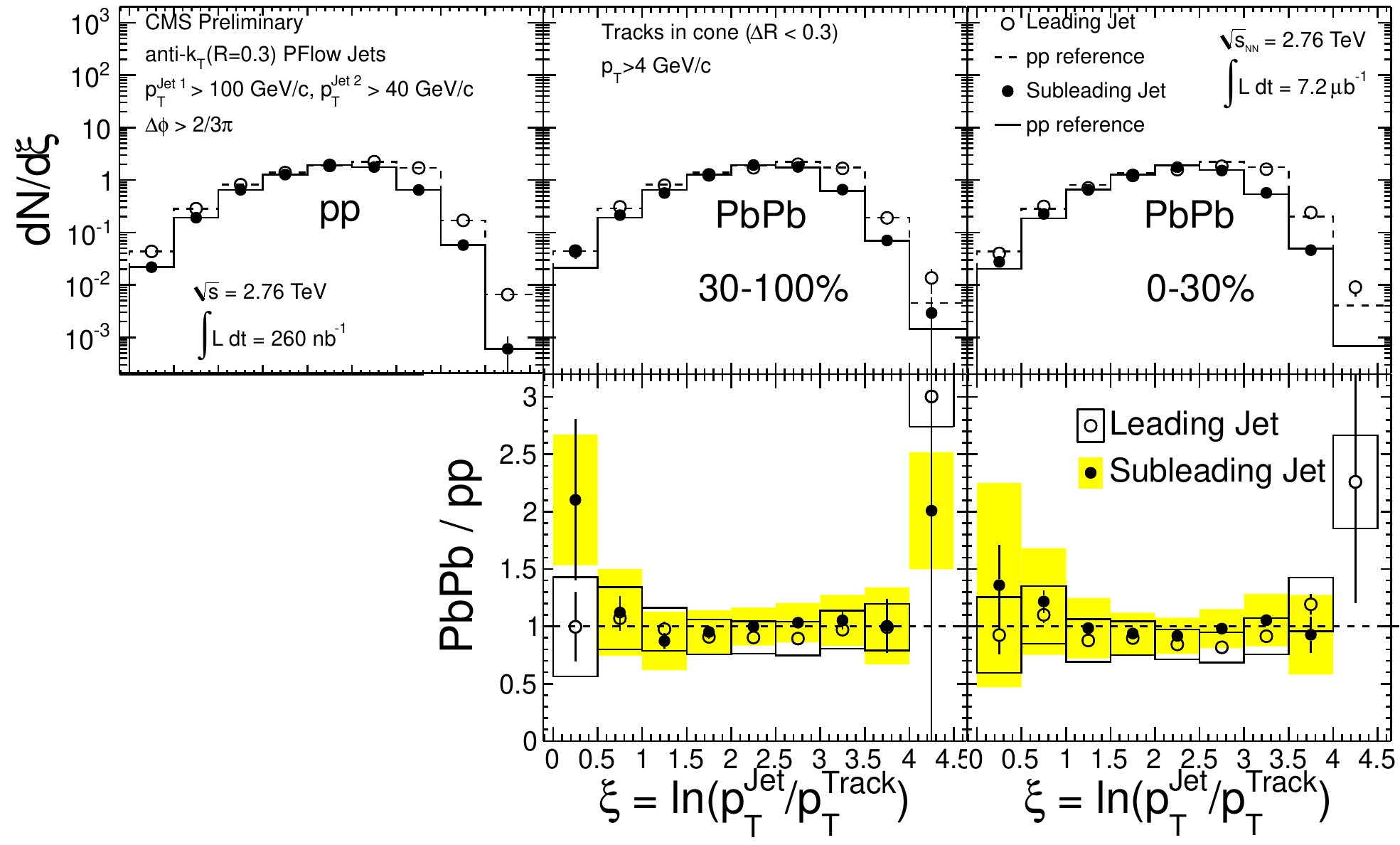}
       \caption{Fragmentation functions for jets produced in pp and in both peripheral (30-100\%) and central (0-30\%) PbPb collisions.}
\label{fig:jet_fragmentation}
\end{center}
\end{figure}

Studying the modification of jets as they are formed from high $\PT$ partons propagating thorough the hot and dense QGP 
has been proposed as a particularly useful tool for probing the produced matter's properties.
Events with at least two jets with the leading (sub-leading) jet with $\PT$ of at least 120 (50) GeV/c 
and with an opening angle $\Delta\phi_{12}>2\pi/3$ were selected to study these medium effects. 
The most striking observation is the large, centrality-dependent, imbalance in the energy of the two jets, as measured in the CMS calorimeters (See Fig.~\ref{fig:dijet_imbalance}). While their energies were very different, the two jets were observed to be very close to back-to-back in the azimuthal plane, implying little or no angular scattering of the partons during their traversal of the medium\cite{jetquenchingCMS}. 
To find the `missing energy', the calorimetric measurement was complemented by a detailed study of low $\PT$ charged particles in the tracker and by using missing $\PT$ techniques. The apparent missing energy was found
among the low $\PT$ particles, predominantly with $0.5<\PT<2$ GeV/c, emitted outside of the sub-leading jet cone.
To further study jet properties in the PbPb environment, the hard component of the fragmentation function was compared to the fragmentation of jets produced in pp collisions at the same energy.  The comparison of fragmentation functions for PbPb and pp events for different centralities is shown in Fig.~\ref{fig:jet_fragmentation}. The distribution of charged particle momenta within the jet, normalized to the measured jet energy, is strikingly the same, within uncertainties, to that seen in the equivalent jets energy produced in pp events.

\begin{figure}[ht!]
\begin{center}
   	\includegraphics[width=0.4\textwidth]{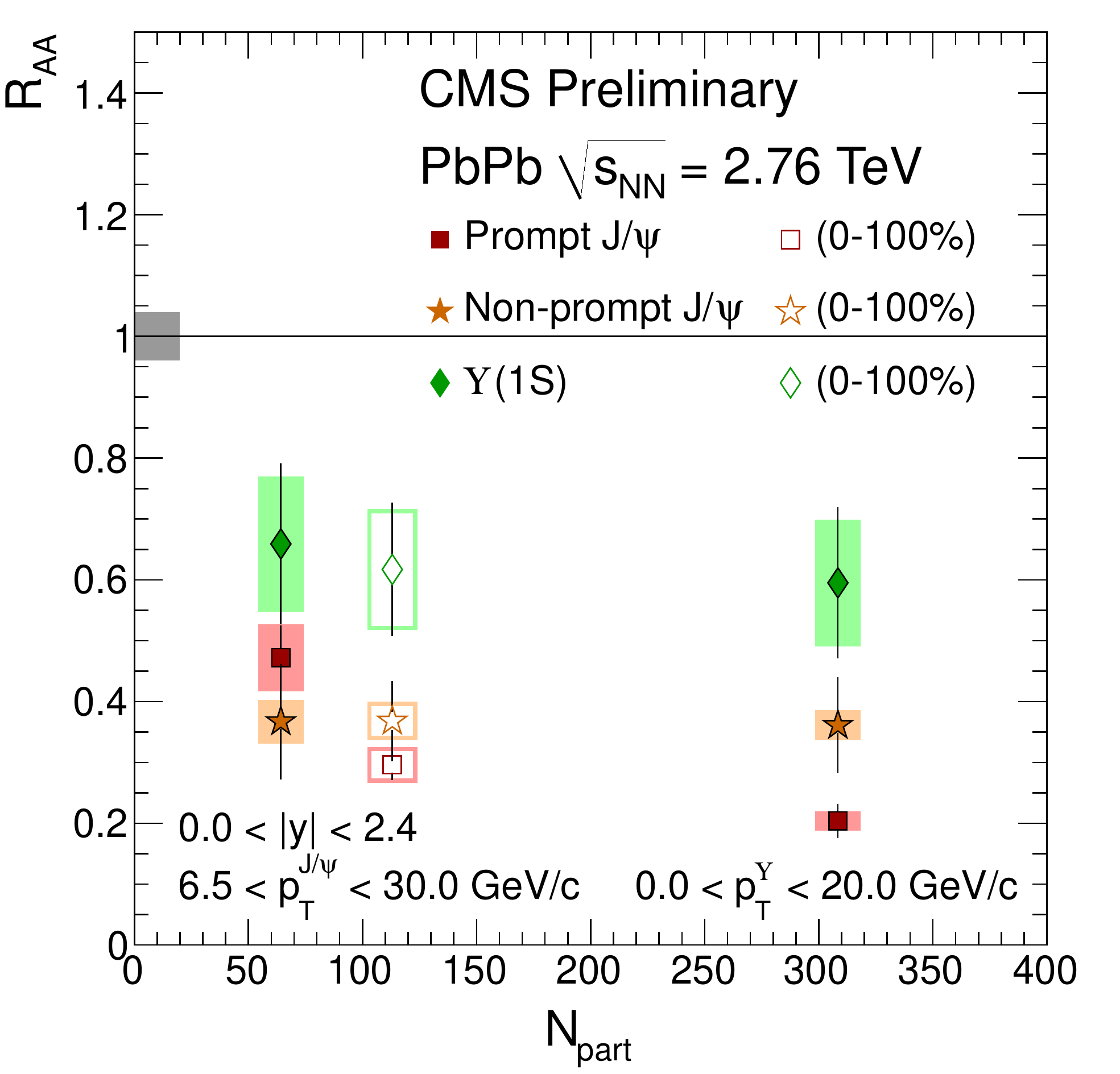}
       \hspace{6mm}
       \includegraphics[width=0.4\textwidth]{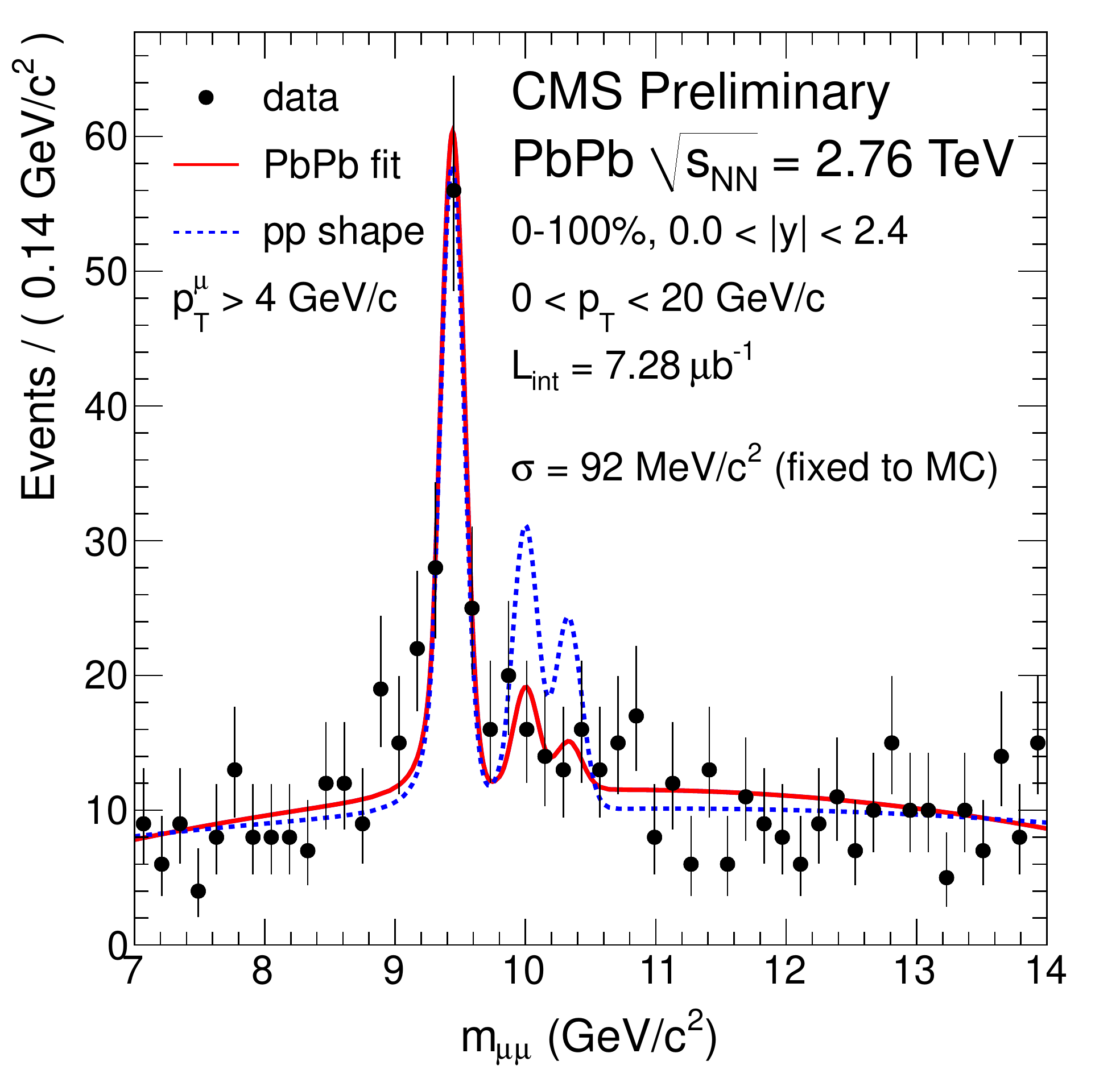}
       \caption{
              (left) Suppression of quarkonia states, $\RAA$ as a function of centrality, for prompt $J/\psi$, non-prompt $J/\psi$ and $\Upsilon(1S)$
       (right) Mass spectrum of $\Upsilon$ family in PbPb collisions compared to the fit obtained in pp events at the same energy.}
\label{fig:quarkonia}
\end{center}
\end{figure}

Quarkonia are important for studying the quark gluon plasma (QGP) since
they are produced early in the collision and their survival is affected by the
surrounding medium. The bound states of charm 
and bottom quarks are expected to be suppressed in heavy-ions, as compared to pp. 
The magnitude of the suppression for different quarkonia states is expected to depend on their binding energy.
By selecting events with opposite-sign dimuons, CMS obtained production rates of $J/\psi$ mesons and of the $\Upsilon$ family. Non-prompt $J/\psi$s (those produced from B-meson decays) could be identified by their displaced decay vertex. The suppressions of prompt and non-prompt $J/\psi$ particles were measured separately. The non-prompt $J/\psi$ suppression is one measure of the quenching of b-quarks. The $R_{AA}$ as function of the number of participants $\Npart$  indicates that high $\PT$ $J/\psi$s are strongly suppressed as shown in Fig.~\ref{fig:quarkonia}(left). The excellent dimuon mass resolution allowed good separation of the three bound states of the $\Upsilon$ family. Fig.~\ref{fig:quarkonia}(right) shows that the excited states, $\Upsilon(2S)$ and  $\Upsilon(3S)$, are suppressed as compared to the $\Upsilon(1S)$. This is compatible with differential melting of quarkonia states in the high temperatures produced by PbPb collisions.

The high energy collisions produced at the LHC allowed for the very first time to observe weak bosons, $Z$ and $W$, in heavy-ion collisions. These weak bosons do not interact with the
strongly interacting medium and, as such, they provide a good reference for the effective parton luminosity. Their cross section is expected to scale only with the nuclear collision geometry. $Z$ decays into pairs of both muons and electrons as well as decays of $W$ to a muon and a neutrino were observed. The $Z$ yield was compared to a binary scaling of the pp cross section and was found to be consistent, within uncertainties, with the expectation of no suppression in the final state.

The CMS collaboration has measured properties of heavy-ion collisions at the highest energies available to-date. Measurements of charged multiplicity, azimuthal asymmetry, dihadron correlations, photons, jets, quarkonia and weak bosons were conducted over a wide azimuthal and rapidity range and with high resolution. The overall picture that emerges is that of a strongly interacting medium that can be well described by hydrodynamics. The availability of high $\PT$ probes allows us
to study how the plasma affects partons with $O(100)$~GeV/c momenta. Large suppression of strongly interacting probes is observed while the photons and weak bosons appear not to be suppressed. The detailed pattern of suppression was measured using quarkonium states with varying binding energies. The scope and the level of detail of all CMS measurements validates the concept of using a general purpose, hermetic detector for studies of heavy-ion collisions.

\section*{References}
\bibliographystyle{unsrt}
\bibliography{bibliography_wyslouch}
\end{document}